\documentclass[a4paper,10pt]{article}

\usepackage{amssymb}
\usepackage{amsmath}
\usepackage[usenames,dvipsnames]{color}
\usepackage{hyperref}
\usepackage[left=1in,right=1in,top=1in,bottom=1in]{geometry}       

\title{Coupling Shape Dynamics to Matter Gives Spacetime}
\author{\bf Henrique Gomes\footnote{\href{mailto:gomes.ha@gmail.com}{gomes.ha@gmail.com}}\\\it University of California at Davis\\ \it One Shields Avenue Davis, CA, 95616, USA \bigskip\\ \bf Tim Koslowski\footnote{\href{mailto:t.a.koslowski@gmail.com}{tkoslowski@perimeterinstitute.ca}}
\\\it Perimeter Institute for Theoretical Physics\\\it 31 Caroline Street, Waterloo, Ontario N2L 2Y5, Canada}

\let\oldmarginpar\marginpar
\renewcommand\marginpar[1]{\oldmarginpar{\color{red}\raggedright\scriptsize #1}}

\newcommand{\mean}[1]{\ensuremath{\lf\langle #1 \rt\rangle }}

\def\be{\begin{equation}}
\def\ee{\end{equation}}
\def\bea{\begin{eqnarray}}
\def\eea{\end{eqnarray}}

\def\lf {\ensuremath{\left}}
\def\rt {\ensuremath{\right}}

\begin{document}

\maketitle

\begin{abstract}
  Shape Dynamics is a metric theory of pure gravity, equivalent to General Relativity, but formulated as a gauge theory of spatial diffeomporphisms and local spatial conformal transformations. In this paper we extend the construction of Shape Dynamics form pure gravity to gravity-matter systems and find that there is no fundamental obstruction for the coupling of gravity to standard matter. We use the matter gravity system to construct a clock and rod model for Shape Dynamics which allows us to recover a spacetime interpretation of Shape Dynamics trajectories. 
\end{abstract}

{\it{ \lq\lq Spacetime is the fairy tale of a classical manifold. It is irreconcilable with quantum effects in gravity and most likely, in a strict sense, it does not exist. But to dismiss a mythical being that has inspired generations just because it does not really exist is foolish. Rather it should be understood together with the story-tellers through whom and in whom the being exist. \rq\rq}}\\ T. Kopf and M. Paschke in \cite{kopf-paschke}.

\section{Introduction}

General Relativity (GR) is a super-theory in the sense that it does not only provide a theory of gravity itself but also provides a general geometric framework for all classical physics. Shape Dynamics (SD) \cite{Gomes:2011zi,Gomes:2010fh,Koslowski:2011jg} is a classical theory equivalent to pure gravity in which the spacetime picture that underlies GR is replaced by an intrinsically spatial picture. In particular the gauge-invariance under spacetime refoliations is traded for a gauge invariance under local spatial conformal transformations that preserve the total spatial volume. It is the purpose of this paper to develop SD further into a framework much like GR. The first step into this direction is to extend SD to a theory that is equivalent to GR coupled to standard matter. This extension of pure SD allows us to gain a spacetime interpretation of SD trajectories by observing how test fields evolve together with pure SD.

It turns out that the construction of a SD extension with standard matter content can be constructed from the appropriate GR theory by straightforward application of the construction principle given in \cite{Gomes:2011zi}. Moreover, gaining a spacetime interpretation of a SD trajectory turns out also to be straightforward  using e.g. a massless free scalar field as a clock and rod model. On the GR side, i.e. from a spacetime point of view, one uses the commutation algebra of vector fields adapted to a given foliation to construct a unique representation for the constraints appearing in the canonical picture \cite{Hojman:1976vp}. The identification of a construction principle for SD is more involved because our present understanding is based on the idea of symmetry doubling. We delegate the motivation and explanation of this idea to future work.

We want to emphasize that our work is very much inspired by the York procedure \cite{York1,York2,York3} to solve the initial value problem in GR, but our construction of SD is by no means equivalent to the York procedure. The two main differences are the following: First, in the York procedure the constant mean curvature condition is treated as a convenient gauge for decoupling the diffeomorphism from the scalar constraint, but the fact that it defines a first class constraint surface and gauge-fixing  properties that enables one to construct a theory\footnote{This construction is only possible if one extends phase space first, and then apply phase space reduction \cite{Gomes:2011zi}.} with local conformal symmetry does not appear in any way in the York procedure\footnote{In fact, the York procedure is only designed as a mathematical device to find valid initial data for GR, in contrast to our approach which deals with the dynamical implementation of conformal symmetry.}. Second, the observation that York scaling can be derived as a canonical transformation on extended phase space using canonical best matching w.r.t. local conformal transformations that preserve the total volume did not appear in work on the initial value problem of GR. However both the volume preserving condition and the property that canonical transformations are morphisms of the Poisson structure are essential to prove the equivalence between GR and SD.    

This paper is organized as follows: We revisit the construction of equivalent gauge theories through the linking theory construction presented in \cite{Gomes:2011zi} in section \ref{sec:preliminaries}. We will use this procedure to construct versions of SD that are equivalent to GR coupled to standard matter. As a warm-up exercise we consider the coupling of a free massless scalar field to gravity in section \ref{sec:warmUp}. This exposes the general algorithm for matter couplings, which is described in section \ref{sec:generalProcedure} and subsequently applied to standard matter in section \ref{sec:standardMatter}, where we consider a class of classical matter theories that includes the standard model. This allows us to consider natural clock and rod models in section \ref{sec:spacetimeRecovery} within SD. These enable us to recover a spacetime interpretation of SD trajectories using the same operational procedure one would use in GR if the spacetime metric was not the dynamical variable of GR to begin with. For a less detailed, more heuristic account of this work, see \cite{Gomes:Loops}. An upcoming account detailing the motivations, the conceptual picture and the general framework in which this work falls is in preparation by the authors. 

\section {Preliminaries}\label{sec:preliminaries}

To make this paper sufficiently self-contained, we review the construction principle for equivalent gauge theories, in particular the construction of Shape Dynamics in vacuum, and then consider the symmetry and locality principle that underlies Shape Dynamics. The following procedure is a summary of that contained in \cite{Gomes:2011zi}. Basically, the procedure in the concrete case of General Relativity consists in first  extending phase space to accommodate extra redundant symmetries obtained through ``conformalization". Then we apply two different particular gauge fixings and follow the Dirac analysis to obtain two dual theories with distinct symmetry properties: GR and Shape Dynamics. 

\subsection{Construction of Equivalent Gauge Theories}

Let us start by assuming a phase space $\Gamma$ coordinatized by configuration variables $\{q_i\}_{i\in\mathcal I}$ and its canonically conjugate momenta $\{p^i\}_{i\in\mathcal I}$, such that we can write the total Hamiltonian of the system as
\begin{equation}
  H=H_o(q,p)+\lambda_\alpha \chi_1^\alpha(q,p)+\lambda^\mu \chi_2^\mu(q,p),
\end{equation}
where $H_o(p,q)$ denotes the primitive Hamiltonian, $\{\chi_1^\alpha\}_{\alpha\in\mathcal A}$ and $\{\chi_2^\mu\}_{\mu\in\mathcal M}$ denote two independent sets of first class constraints and $\lambda$ denote Lagrange multipliers. We now consider $\Gamma$ as a trivially embedded subspace of a larger phase space $\Gamma_{ext}=\Gamma\times\Gamma_\phi$, where $\Gamma_\phi$ can be coordinatized by auxiliary configuration variables $\{\phi_\alpha\}_{\alpha\in\mathcal A}$ and their canonically conjugate momenta $\{\pi^\alpha\}_{\alpha\in\mathcal A}$. This implies that all phase space functions on $\Gamma$ strongly Poisson commute with $\pi^\alpha$, so we can add these consistently to the first class constraints. Now we apply the canonical transformation in extended phase space              
\begin{equation}
 \begin{array}{rclcrcl}
   q_i&\to&Q_i(q,\phi)&,& p^i&\to&P^i:=\left(M^{-1}\right)^i_j p^j\\
   \phi_\alpha&\to&\phi_\alpha&,&\pi^\alpha&\to&\Pi^\alpha:=\pi^\alpha-R^\alpha_j \left(M^{-1}\right)^j_k p^k.
 \end{array}
\end{equation}
This transformation is part of the canonical treatment Barbour's ''best matching``, but we will not dwell on this point here.
We use the shorthand the shorthand $M^i_j=\frac{\partial Q_j}{\partial q_i}=\frac{\partial \dot Q_j}{\partial \dot q_i}$ as well as $R^\alpha_j:=\frac{\partial Q_j}{\partial \phi_\alpha}=\frac{\partial \dot Q_j}{\partial \dot\phi_\alpha}$.

We thus have a gauge theory with Hamiltonian
 \begin{equation}
  H=H_o(Q,P)+\lambda_\alpha \chi_1^\alpha(Q,P)+\lambda^\mu \chi_2^\mu(Q,P)+\lambda_\beta\Pi^\beta,
\end{equation}
 where we have extended the first class constraints by $\Pi^\beta$. This reduces to the original gauge theory by imposing the gauge-fixing conditions $\phi_\alpha=0$ and performing the phase space reduction by replacing $(q,p,\phi,\pi)\to(q,p,0,0)$.

 If the constraint system $\chi^\alpha_1(Q(q,\phi),P(q,\phi,p))=0$ can be solved for the $\phi_\alpha=\phi^o_\alpha(q,p)$ then one can impose the gauge fixing condition $\pi^\alpha$ and perform the phase space reduction through the replacement $(q,p,\phi,\pi) \to (q,p,\phi^o(q,p),0)$, which trades the first class constraints $\chi^\alpha$ for $R^\alpha_j \left(M^{-1}\right)^j_k p^k$ and primitive Hamiltonian $H_o(Q(q,\phi^o(q,p)),P((q,\phi^o(q,p)),p) )$. This theory lives on the same phase space $\Gamma$ as the original theory and possesses (under particular gauge fixings) an equivalent initial value problem and equivalent equations of motions as the theory we started out with. The only difference is a different set of gauge symmetries.

\subsection{Pure Shape Dynamics}

The procedure described in the previous subsection was applied in \cite{Gomes:2011zi} to construct Shape Dynamics as a theory equivalent to ADM gravity on a compact Cauchy surface $\Sigma$ without boundary. For this we start with the standard ADM phase space $\Gamma_{ADM}=\{(g,\pi):g\in \mathrm{Riem},\pi\in T_g^*(\mathrm{Riem})\}$, where $\mathrm{Riem}$ denotes the set of Riemannian metrics on $\Sigma$, that satisfy a sufficient differentiability condition, and the usual first class ADM constraints (i.e. the scalar constraints $S(N)$ and momentum constraints $H(\xi)$) thereon. We extend the ADM phase space with the phase space of a scalar field $\phi(x)$ and its canonically conjugate momentum density $\pi_\phi(x)$, which we introduce as additional first class constraints $\mathcal{Q}(x)=\pi_\phi(x)\approx 0$, and use the canonical transformation $T_\phi$ generated by the generating functional $F=\int d^3x\left(g_{ab}e^{4\hat \phi}\Pi^{ab}+\phi\Pi_\phi\right)$, where $\hat \phi(x):=\phi(x)-\frac 1 6 \ln\langle e^{6\phi}\rangle_g$ using the mean $\langle f\rangle_g:=\frac 1 V \int d^3x\sqrt{|g|} f(x)$ and 3-volume $V_g:=\int d^3x\sqrt{|g|}$. This yields the first class set of constraints
\begin{equation}
 \begin{array}{rcl}
   T_\phi S(N)&=&T_\phi\left(\frac{\pi^{ab}\pi_{ab}-\frac{1}{2}\pi^2}{\sqrt g}-\sqrt g R\right)\\
   T_\phi H(\xi)&\approx&\int d^3x \pi^{ab}\mathcal L_\xi g_{ab}+\pi_\phi\mathcal{L}_\xi\phi\\
   T_\phi Q(\rho)&=&\pi_\phi-4(\pi-\langle\pi\rangle\sqrt g)
 \end{array}
\end{equation}
$Q(1)$ is not a constraint, since $\langle\pi_\phi\rangle:=\int d^3 x \pi_\phi$ Poisson commutes with $\hat \phi$ and $T_\phi$ is a canonical transformation. This allows us to perform the phase space reduction for the gauge fixing condition $\pi_\phi(x)=0$. The end result is the following set of first class constraints, which define Shape Dynamics
\begin{equation}\label{equ:pureSDconstraints}
 \begin{array}{rcl}
   H_{SD}&=&T_{\phi_o} S(N_o)\\
   H(\xi)&=&\int d^3x \pi^{ab}\mathcal{L}_\xi g_{ab}\\
   Q(\rho)&=&\int d^3x \rho\left(\pi-\langle \pi\rangle\sqrt{|g|}\right),
 \end{array}
\end{equation}
where $\phi_o(g,\pi)$ is such that $H_{SD}=0$ combined with $\phi=\phi_o(g,\pi)$ is equivalent to $T_\phi S(x)=0$ (at the surface $\pi_\phi=0$). Moreover, $N_o$ denotes the CMC lapse function, $\pi=\pi^{ab}g_{ab}$ and $\langle\pi\rangle=\frac 1 V\int d^3 x\pi$. The constraints $H(\xi)$ generate spatial diffeomorphisms and the constraints $Q(\rho)$ generate conformal transformations that leave the total spatial volume invariant.

\section{Warm-Up: Free Scalar Field}\label{sec:warmUp}

The most natural way to approach the coupling of a scalar field, is not to try to develop it directly in SD, but to make use of the linking theory procedure to facilitate its introduction. We shall concentrate on the case where we have a closed spatial manifold without boundary.

Let us define the fields we shall be working with. The usual Hamiltonian density for a scalar field $\psi$,  is given by
\be H_\psi =\frac{\pi_\psi^2}{\sqrt g}+g^{ab}\nabla_a\psi\nabla\psi_b\sqrt g
\ee
where $\pi_\psi$ denotes the conjugate momentum density (canonically conjugate to $\psi$) to the scalar field. The original gravitational constraints amended by the constraints arising from the coupling to the scalar field can be written as
\begin{equation}\label{equ:scalar_fieldHamiltonian}
 \begin{array}{rcl}
   S(N)&=&\int d^3x N\left(\frac{\pi^{ab}\pi_{ab}-\frac{1}{2}\pi^2+\pi_\psi^2}{\sqrt g}-\sqrt g (R-g^{ab}\nabla_a\psi\nabla\psi_b)\right)\\
   H^a(\xi_a)&=&\int d^3 x(g_{ab}\mathcal{L}_\xi\pi^{ab}+\psi \mathcal{L}_\xi\pi_\psi)
  \end{array}
  \end{equation}
where for ease of manipulation we wrote the smeared version of the diffeomorphism constraint.

We now embed the original system into an extended phase space including the auxiliary variables $(\hat\phi,\pi_{\hat\phi})$.
The nontrivial canonical Poisson brackets are
\begin{equation}\label{equ:canonical_relations}
  \begin{array}{rcl}
    \{g_{ab}(x),\pi^{cd}(y)\}&=&\delta^{(cd)}_{ab}\delta(x,y)\\
    \{\phi(x),\pi_\phi(y)\}&=&\delta(x,y)\\
    \{\psi(x),\pi_\psi(y)\}&=&\delta(x,y).
  \end{array}
\end{equation}
The extended phase space for these fields is now given by:
$$(g_{ij},\pi^{ij}, \psi, \pi_\psi,\phi,\pi_\phi)\in\Gamma_{\mbox{\tiny {Ex}}}:=\Gamma_{\mbox{\tiny {Grav}}}\times \Gamma_{\mbox{\tiny {Scalar}}}\times \Gamma_{{\mbox{\tiny {Conf}}}}$$
with the additional constraint :
\begin{equation}\pi_\phi\approx 0
\end{equation}

Following \cite{Gomes:2011zi}, we construct the generating function \begin{equation}\label{equ:generatingFunctional2}F_\phi:=\int_\Sigma d^3x( g_{ab}(x)e^{4\hat\phi(x)}\Pi^{ab}(x)+\phi\Pi_\phi+\psi\Pi_\psi),\end{equation} we find the canonical transformation:
\begin{equation}\label{equ:transformations}
 \begin{array}{rcl}
   g_{ab}(x)&\to& T_\phi g_{ab}(x)=e^{4\hat\phi(x)}g_{ab}(x)\\
   \pi^{ab}(x)&\to&T_\phi \pi^{ab}(x):=\Pi^{ab}(x)=e^{-4\hat\phi}(\pi^{ab}-\frac{g^{ab}}{3}\sqrt {g}\langle \pi\rangle (1-e^{6\hat\phi}))\\
    \phi(x)&\to&T_\phi \phi(x)=\phi(x)\\
   \pi_\phi(x)&\to&T_\phi \pi_\phi(x):=\Pi_\phi=\pi_\phi-4(\pi-\langle \pi\rangle\sqrt g)\\
      \psi(x)&\to&T_\phi \psi(x)=\psi(x)\\
   \pi_\psi(x)&\to&T_\phi \pi_\psi(x):=\Pi_\psi=\pi_\psi
 \end{array}
\end{equation}
and subsequently use these transformed variables to construct three sets of constraints: the transformed scalar- and diffeomorphism-constraint of GR  as well as the transform of $\pi_\phi$,
\begin{equation}
 T_\phi S;~ T_\phi H^a ;~
  \mathcal{Q}:=\pi_\phi-4(\pi-\langle \pi\rangle\sqrt g)
\end{equation}

 The linking theory gravitational Hamiltonian is:
\be H_{\mbox{\tiny {Total}}}=\int d^3x [N(x){\mathcal {T}}_\phi S(x)+\xi^a(x){\mathcal {T}}_\phi H_a(x)+\rho(x)\mathcal{Q}(x)]
\ee Now we use the gauge fixing $\pi_\phi=0$, and find that it weakly commutes with all constraints except for ${\mathcal {T}}_\phi S$.
We get for the propagation of the gauge fixing condition $\pi_\phi=0$:
\be\label{equ:gfPB} \{T_\phi S(N), \pi_\phi(x)\}=T_\phi\left[-\frac{3}{2}S(x)+2( \Delta_{\mbox{\tiny scalar}}N(x)-\mean{ \Delta_{\mbox{\tiny scalar}}N})\right]
\ee
where we have defined the second order elliptic differential operator 
\be\label{equ:def:Delta_scalar}\Delta_{\mbox{\tiny scalar}}:= \nabla^2-\frac{\pi\langle\pi\rangle}{4\sqrt g}-R+g^{ab}\nabla_a\psi\nabla_b\psi\ee
By the homogeneous  form of  equation \eqref{equ:gfPB} (and as we have already evaluated the Poisson bracket), evaluation on the constraint surface defined by $T_\phi S(x)=0$ and $\mathcal{Q}(x)=0$, is equivalent to the $\phi$-transform of the evaluation of the term in brackets on the constraint surface defined by $S(x)=0$ and $D(x)=0$. 

We now note that the function 
 $ -\frac{\pi\langle\pi\rangle}{4\sqrt g}-R+g^{ab}\nabla_a\psi\nabla_b\psi$ is non-positive on the constraint surface. To do so, we use the scalar constraint to get the operator  $\Delta_{\mbox{\tiny scalar}}$  in the form of: 
\be -\frac{G_{abcd}\pi^{ab}\pi^{cd}-\pi_\psi^2}{\sqrt g}-\frac{1}{4}\pi\mean{\pi}+\sqrt g\nabla^2\approx \sqrt{g}(\nabla^2-\frac{1}{12}\langle\pi\rangle^2)-\frac{\sigma^{ab}\sigma_{ab}+\pi_\psi^2}{\sqrt g}
\ee
where $\sigma^{ab}$ is the traceless part of the momenta and we also used the constraint $\mathcal{Q}$.
We are thus guaranteed by The Maximum Principle for second order elliptic operators  that  the Green's function for $ \Delta_{\mbox{\tiny scalar}}N$ exist and are unique. Finally, this means there exists a unique  solution $N_0[g,\pi,\pi_\psi]$ for the equation 
$\Delta_{\mbox{\tiny scalar}}N=c$ where $c$ is a given spatial constant, and such that $\mean{N_0}=1$.

We can then define the gauge-fixed part of ${\mathcal {T}}_\phi S$ as $\widetilde {{\mathcal {T}}_\phi S}={\mathcal {T}}_\phi S-{\mathcal {T}}_\phi S(N_0)$. It can be now shown that 
$$ \{  \widetilde {{\mathcal {T}}_\phi S}, \pi_\phi\}= \frac{\partial  \widetilde {{\mathcal {T}}_\phi S}}{\partial \phi}
$$ is invertible and that we can use the implicit function theorem to 
write  $\widetilde {{\mathcal {T}}_\phi S}$ as $\phi-\phi_0(g,\pi,\psi,\pi_\psi)$ \cite{Gomes:2011zi}. This constraint exhausts the gauge fixing $\pi_\phi=0$ (they have invertible Poisson bracket), and, as second class constraint, it can be set strongly to zero alongside $\pi_\phi$ to eliminate the extra variables and allow us to use the usual Poisson bracket instead of the Dirac ones.

The two outstanding features of the coupling are that the constraint $\mathcal{Q}$ does not depend on the scalar field $\psi$ and that one can still uniquely solve the lapse fixing equation for a functional $N_0[g,\pi,\pi_\psi]$ such that $\langle N_0\rangle=1$. We thus have well defined shape dynamics coupled to a scalar field given by the first class constraints:
\be\label{equ:scalar_field_constraints} \langle {\mathcal {T}}_{\phi_0} S N_0\rangle;~~\{ H^a(x) ,~ x\in \Sigma\};~~\{D(x):=4(\pi(x)-\langle \pi\rangle\sqrt g(x)),~x\in\Sigma\}
\ee where we have used that with the reduction ${\mathcal {T}}_\phi H^a\rightarrow H^a$.

\section{General Procedure}\label{sec:generalProcedure}

The treatment of the free scalar field in the previous section exposed the explicit procedure that allows one to construct SD matter systems. Let us now describe this procedure before applying it to a concrete model in the next section.

\subsection{What constrains the coupling to the conformal factor?}

When we try to couple different fields to gravity and construct a linking theory, we have to answer one important question: How do we have to scale the new fields? i.e. what exponent is viable in the conformal coupling of a scalar field,    $\psi\rightarrow e^{\alpha\hat\phi}\psi$? This is an important issue because if the scaling is not correct we could encounter two difficult obstructions.

The first obstruction is that if we are dealing with a field that possesses some kind of gauge symmetry it might not be possible to find a constraint $\mathcal{Q}$  that is first class with respect to the gauge constraint. The second is that the conserved charge (which we call $D$) implicit inside $\mathcal{Q}$  that defines the foliation  might depend on the field. In this case, there might be an even worse consequence if the field possesses some sort of gauge symmetry (like electromagnetism). For then the charge could turn out to depend on the gauge potential. This could be the case even if we can find a field-dependent $D$ that is first class with respect to the gauge generator of the field. Indeed with any other choice of scaling than the one chosen in the text, this is what happens with any Yang-Mills type interaction, where the generator of the gauge symmetry is the Gauss constraint.
 
What we have learned from the previous section is that the coupling of the scalar field \emph{does}  work for the particular choice of \emph{neutral} scaling $\alpha=0$, i.e. non-gravitational fields have conformal weight zero. This means that fields are only scaled in the sense that they are ``carried along" by the scaling of the spatial metric.

We will not show it here, but in the Yang-Mills case it is possible to make a scaling choice different from this one  and still such that the Gauss constraint is propagated. This choice however introduces an anomaly in the sense that the $D$ constraint (which defines the foliation in the space-time picture) becomes dependent on the gauge choice of the Yang-Mills field. So let us carry onwards and apply this ``neutral coupling"  in the general case.

 \subsection{Explicit coupling}

We follow the construction principle for pure Shape Dynamics and suppose we have a gravity-matter system on a compact manifold without boundary with constraints
\begin{equation}
 \begin{array}{rcl}
   H(\xi)&=&\int d^3x \left( \pi^{ab} (\mathcal L_\xi g)_{ab}+ \pi^A \mathcal (L_\xi \phi)_A \right)\\
   S(N)&=&\int d^3x\left(\frac 1{\sqrt{|g|}}\pi^{ab}G_{abcd}\pi^{cd}-(R-2\Lambda)\sqrt{|g|}+H_{\mbox{\tiny matter}}(g_{ab},\phi_A,\pi^A)\right)N(x)\\
G^\alpha(\lambda_\alpha)&=&\int d^3x G^\alpha(g_{ab},\phi_A)\lambda_\alpha\sqrt g
 \end{array}
\end{equation}
where we denote the non-gravitational configuration degrees of freedom collectively by $\phi_A$ and their canonically conjugate momenta $\pi^A$. We assume that the matter Hamiltonian $H_{\mbox{\tiny matter}}$ does neither contain $\pi^{ab}$ nor any spatial derivatives of $g_{ab}$, which seems to be an assumption that is realized in nature. We furthermore assume the constraint associated to internal gauge symmetries to be a functional of only the ``position" variables $\phi^A, g_{ab}$. 

To construct a linking theory we extend phase space by a scalar $\phi(x)$ with canonically conjugate momentum density $\pi_\phi(x)$ and impose the additional first class constraint
\begin{equation}
 Q(\rho)=\int d^3x \rho(x) \pi_\phi(x),
\end{equation}
which eliminates the phase space extension on shell. To perform canonical best matching for volume preserving conformal transformations, we have to specify the conformal weight of matter fields. As we have mentioned before, from practically working with explicit matter Hamiltonians, it seems as if vanishing conformal weight for all matter fields is the preferred choice to simultaneously achieve decoupling of gauge constraints and retaining a unique global Hamiltonian constraint, but there may be exceptions to this scenario. We thus use the generator
\begin{equation}
 F=\int d^3x\left(e^{4\hat \phi} g_{ab}\Pi^{ab}+\phi \Pi_\phi+\phi_A \Pi^A\right)
\end{equation}
to implement the conformal Kretschmannization. The resulting canonical transformation is given by  equation \eqref{equ:transformations} with 
\begin{equation}\label{equ:new_transformations}
 \begin{array}{rcl}
   \phi_A&\to&\phi_A\\
   \pi^A&\to&\pi^A,
 \end{array}
\end{equation}
so the constraints $Q(\rho)$ respectively $H(\xi)$ transform weakly into
\begin{equation}
 \begin{array}{rcl}
   Q(\rho)&=&\int d^3x \rho\left(\pi_\phi-4\left(\pi-\langle\pi\rangle\sqrt{|g|}\right)\right)\\
   H(\xi)&=&\int d^3x \left(\pi^{ab}\mathcal (L_\xi g)_{ab} +\pi_\phi \mathcal L_\xi \phi + \pi^A(\mathcal L_\xi \phi)_A\right).
  \end{array}  
\end{equation}
In the cases of interest that we have studied, as we will see shortly, the gauge constraint decouples from the conformal transformation, so that $T_\phi G^\alpha\propto G^\alpha\approx 0$.

After imposing the best-matching condition $\pi_\phi(x)=0$ we always get an equation of the form of \eqref{equ:gfPB}:
\be  \{T_\phi S(N), \pi_\phi(x)\}=T_\phi\left[-\frac{3}{2}S(x)+2( \Delta_{\mbox{\tiny matter}}N(x)-\mean{ \Delta_{\mbox{\tiny matter}}N})\right]
\ee
In the general case one has to compute explicitly the entire bracket $\{ T_\phi H_{\mbox{\tiny matter}}, \pi_\phi\} $ and work out the appropriate form of the $\Delta_{\mbox{\tiny matter}}$ to check its invertibility. Let us here restrict to the simpler case occurring when the matter Hamiltonian does not contain spatial derivatives of the metric tensor nor metric momenta. In this case one gets:
\be\label{equ:def:Delta_gen}\Delta_{\mbox{\tiny matter}}:= \nabla^2-\frac{\pi\langle\pi\rangle}{4\sqrt g}-R+\frac{1}{2\sqrt{|g|}}\left(\frac{\delta H_{\mbox{\tiny matter}}}{\delta g_{ab}}g_{ab}+\frac{3}{2}H_{\mbox{\tiny matter}}\right)\ee
On the constraint surface $T_\phi S=0$ and $\mathcal{Q}=0$, the end result is equivalent to taking
\be \Delta_{\mbox{\tiny matter}}\approx (\nabla^2-\frac{1}{12}\langle\pi\rangle^2)-\frac{\sigma^{ab}\sigma_{ab}}{g} 
+\frac{1}{2\sqrt{g}}\left(\frac{\delta H_{\mbox{\tiny matter}}}{\delta g_{ab}}g_{ab}-\frac{1}{2}H_{\mbox{\tiny matter}}\right)\ee
Thus in this case the criterium for invertibility of the operator rests on:
\be\label{equ:inv_criterium}\frac{1}{2\sqrt{g}}\left( \frac{\delta H_{\mbox{\tiny matter}}}{\delta g_{ab}}g_{ab}-\frac{1}{2}H_{\mbox{\tiny matter}}\right)\leq  \sqrt{g} \frac{1}{12}\langle\pi\rangle^2 +\frac{\sigma^{ab}\sigma_{ab}}{ g} 
\ee
in particular if 
\be\label{equ:bound0} \frac{\delta H_{\mbox{\tiny matter}}}{\delta g_{ab}}g_{ab}-\frac{1}{2}H_{\mbox{\tiny matter}}\leq 0\ee
under the previously assumed conditions  the field can always be included in our model.  In some cases the inequality \eqref{equ:inv_criterium} cannot be attained, in others it implies a bound on the density of the fields, and yet in others it is always obeyed, since \eqref{equ:bound0} is  valid. Examples of the first are obtained from fields with four-dimensional conformal coupling, examples of the second kind are obtained by adding a mass potential term to the Hamiltonian, of the form $\psi^2\sqrt g$, and of the third are Yang-Mills and the massless scalar obtained in the previous session. Another formula that might be useful is to rewrite \eqref{equ:bound0} for the specific case of homogeneous scaling. If $H_{\mbox{\tiny matter}}$ scales with power $n$, we have the equivalent condition (assuming $H_{\mbox{\tiny matter}}$ is positive):
\be\label{equ:hom_scal} n-\frac{1}{2}\leq 0.
\ee

\section{Standard Matter}\label{sec:standardMatter}

The standard model, as well as some interesting extensions, are Yang-Mills gauge theories that are minimally coupled to chiral fermions, supplemented with a Higgs multiplet to furnish spontaneous symmetry breaking. We now investigate a model of this type. Let us denote the Yang Mills fields by $A$, the fermion fields by $\psi$ and the Higgs fields by $\varphi$ and consider a gravity-matter action of the form:
\begin{equation}
 S=\int d^4x \left(\sqrt{|g|} R-2\Lambda\sqrt{|g|} + L_{\mbox{\tiny YM}}(g,A) + L_{\mbox{\tiny Dirac}}(g,A,\psi) + L_{\mbox{\tiny Higgs}}(g,A,\varphi)\right),
\end{equation}
where only in this equation $g$ denotes the spacetime metric; subsequently it will again denote the spatial metric. We will not go through the cumbersome Legendre transform, but rather present the canonical picture right away. For now the only restriction is that we assume a globally hyperbolic spacetime. Before we write down the  model, we need to introduce some notation.

In the spacetime description one keeps time-components of the Yang-Mills potential to retain manifest 4-covariance. A minimal Hamiltonian description does not need these redundant fields as their role is reduced to that of Lagrange multipliers. We will thus only need the spatial components of the gauge potential,  denoted by $A_a^C dx^a \tau_C$, and its canonically conjugate momentum density, $E^a_C\partial_a\tau^C$, where capital indices label the components of the gauge Lie-algebra whose generators are denoted by $\tau$. We denote the components of the Higgs field by $\varphi^\alpha$ and the canonically conjugate momentum density by $\pi_{\varphi\alpha}$. We denote the structure constants of the gauge group by $C$ and the representation of its Lie algebra on particle multiples by $T$. We will use density weight $\frac 1 2$ spinors $\psi$ (with adjoint $\bar \psi$) and suppress multiplet indices. 

Our work from now on technically parallels the treatment of the initial value problem in the York procedure \cite{York3,IsenbergNester1,IsenbergNester2}. The models under consideration possess the constraints:
\begin{equation}
 \begin{array}{rcl}
   G(\lambda)&=&\int d^3x \,\lambda^B \left( \partial_aE^a_B+T^A_{BC}E^a_A A_a^C + \pi_{\varphi \alpha} T^\alpha_{B\beta} \varphi^\beta + \textrm{more} \right)\\
   H(\xi)&=&\int d^3x \left(\pi^{ab}(\mathcal L_\xi g)_{ab}+E^a_B (\mathcal L_\xi A)^B_a+\pi_{\varphi\alpha} \mathcal L_{\xi} \varphi^\alpha+\textrm{more}\right)
  \end{array}
\end{equation}
and the refoliation constraints
\begin{equation}
 \begin{array}{rcl}
   S(N)&=&\int d^3x\, N \left(S_{\mbox{\tiny Grav.}}+S_{\mbox{\tiny YM}}+S_{\mbox{\tiny Higgs}}+\textrm{more}\right)\\
   S_{\mbox{\tiny YM}}(x)&=&\frac 1 2 \frac{g_{ab}}{\sqrt{|g|}}E^a_BE^b_B + \frac 1 4 \sqrt{|g|}g^{ac}g^{bd}F^B_{ab}F^B_{cd}\\
   S_{\mbox{\tiny Higgs}}(x)&=&\frac 1{2\sqrt{|g|}} \pi_{\varphi\alpha}^2+\frac 1 2 \sqrt{|g|}g^{ab}(D_a\varphi)^\alpha(D_b\varphi)^\alpha+V(\varphi)\sqrt{|g|},
 \end{array}
\end{equation}
where $S_{\mbox{\tiny Grav.}}$ denotes the pure gravity scalar constraint and $D_a$ denotes the gravity-Yang-Mills covariant derivative on the Higgs bundle. The term ''more`` is retained for generality, in particular in the Fermi sector, where we intentionally omit a specific form (mixing, masses, Dirac vs. Majorana etc) to avoid distraction from the main point of the paper\footnote{A proper treatment of couplings to Dirac fields requires the introduction of a spin structure, which is the subject of forthcoming work. If we where just interested in the initial value problem we could proceed as  \cite{York3} with the modification that we use canonical spinors of density weight $\frac 1 2$.}. It turns out that the only essential ingredient is the density weight $\frac 1 2$, so usual matter contributions do not couple to derivatives of the metric. In addition there will in general be helicity constraints, which are however algebraic, so the helicity constraints before and after a conformal transformation are equivalent. We can thus also put helicity constraints aside as they do not impinge in our construction. 

Let us now construct the SD theory that is equivalent to GR with such matter couplings. We extend phase space to include $\phi(x)$ and its canonically conjugate momentum $\pi_\phi(x)$ and choose ``neutral coupling`` for the matter. This is to say that we use a canonical transformaiton resulting in: 
\begin{equation}
 \begin{array}{rclcrcl}
   g_{ab}&\to& e^{4 \hat\phi}g_{ab}&&\pi^{ab}&\to&e^{-\hat \phi}\left(\pi^{ab}-\langle \pi \rangle (1-e^{\hat \phi})\sqrt{|g|}g^{ab}\right)\\
   \phi&\to&\phi&&\pi_\phi&\to&\pi_\phi-4\left(\pi-\langle\pi\rangle\sqrt{|g|}\right)\\
   A_a^I&\to&A_a^I&&E^a_I&\to&E^a_I\\
   \varphi^\alpha&\to&\varphi^\alpha&&\pi_{\varphi \alpha}&\to&\pi_{\varphi \alpha}\\
   \psi & \to & \psi && \psi^* &\to& \psi^*.
 \end{array}
\end{equation}
We see immediately that the Gauss- and diffeomorphism- constraints are weakly preserved for ``neutral coupling``. None of the matter contributions to the scalar constraints contain derivative couplings, so we can use equation (\ref{equ:inv_criterium}) to determine whether the partial gauge fixing $\pi_\phi(x)=0$ propagates. For this it is advantageous to rearrange the terms in the matter contributions according to their homogeneous conformal scaling
\begin{equation}
 \begin{array}{rcl}
  S_{-3/2}&=&\frac 1 2\frac{\pi_{\varphi\alpha}^2}{\sqrt{|g|}}\\
  S_{-1/2}&=&\frac 1 2 \frac{g_{ab}}{\sqrt{|g|}}E^a_AE^b_A+\frac 1 4 \sqrt{|g|}g^{ac}g^{bd}F^A_{ab}F^A_{cd}\\
  S_{+1/2}&=&\frac 1 2 \sqrt{|g|}g^{ab}(D_a\varphi)^\alpha(D_b\varphi)^\alpha\\
  S_{+3/2}&=&\sqrt{|g|}V(\varphi),
 \end{array}
\end{equation}
where we did not explicitly denote Fermi contributions. The only term that does not immediately obey the bound \eqref{equ:hom_scal} is  $S_{+3/2}$, which  may imply that the modulus of the mean extrinsic curvature $|\langle\pi\rangle|$ is bounded from below, or equivalently, that the potential $V(\varphi)$ is bounded from above.

 However, as soon as this restriction is imposed, one can use the implicit function theorem to prove the existence of a functional $\phi_o$ that solves $\tilde T_{\phi_o}S(x)=0$. Defining $H_{SD}=T_{\phi_o}S(N_o)$ completes the construction of SD for this matter model. 

We should note that, except for the Higgs field,  the entire Standard Model (writing the coupling of the fermions using inverse triads) couples with the same weight: $-1/2$. Beyond the Standard Model, but still within the realm of observed phenomena, the cosmological constant also stands out as being constrained by a bound in equation \eqref{equ:inv_criterium}. Within the Standard Model, the only other component which implies a bound is the Higgs potential, $ S_{+3/2}$.

We here very briefly mention a very real violation of the bound. Take
the de Sitter solution to Einstein's equations, which in global coordinates has a metric of the form 
$$ds^2=-d\tau^2+\alpha 2\cosh^2(\tau/\alpha)d\Omega^2
$$
where $d\Omega^2$ is the line element for the 3-sphere and $\alpha$ is the de Sitter length, related to the cosmological constant as $\Lambda=3/\alpha^2$. In these coordinates, de Sitter is conveniently in a CMC foliation, and starts out as an infinitely large sphere which contracts to a minimum radius and expands again to infinite radius. If one performs the usual Legendre transformation to this solution, one can see that it does not satisfy the bound \eqref{equ:inv_criterium}.  In this paper we do not purport to interpret the effect that this has for the construction of the theory. It doesn't seem to be a trait of the symmetry content of the space, as one can ruin the symmetry and still not satisfy the bound, in spacetimes arbitrarily close to De Sitter. 

\section{Reconstruction of Spacetime}\label{sec:spacetimeRecovery}

General Relativity is a theory of the spacetime metric, but the physical interpretation of the dynamical metric as the geometry of the universe arises through a clock and rod model given by the matter content of the theory.  We can thus view the operationally defined geometry as fundamental and only accept it as a nice feature of General Relativity to use the spacetime metric as a fundamental field. Terms like ``light cone`` put the operational meaning of geometry to the forefront. Shape Dynamics does not immediately provide a spacetime metric at a glance, but a spacetime interpretation of Shape Dynamics comes operationally out of a clock and rod model in the same way as it does in General Relativity. The simplest clock and rod model is a multiplet of massless free scalar fields, which we will consider here. 

For this we assume that the field strength $\psi^i(x,t)$ of the $i$ components of a scalar multiplet and canonically conjugate momentum density $\pi^i_\psi(x,t)$ can be prepared at every point $x\in \Sigma$ at a given time $t_o$ and that both are measurable at later times, such the we can at least measure the time derivatives of the field- and momentum-components. Moreover, we assume that the fields can be prepared as test fields, i.e. the field strength and momentum density is small enough, so the backreaction on gravity can be safely neglected. To recover the spacetime metric at a point $(x_o,t_o)$ we choose a local chart with coordinates $(t,x^a)$ and consider the equations of motion generated by the massless Hamiltonian $H=H_{grav.}+\int d^3x \frac N 2 \left(\frac{\pi_{\psi^i}^2}{\sqrt{|g|}}+\sqrt{|g|}g^{ab}\psi^i_{,a}\psi^i_{,b}\right)+\int d^3x \pi_{\psi^i}\mathcal L_\xi \psi^i$ for massless free test fields coupled to gravity:
\begin{equation}
 \begin{array}{rcl}
   \{H,\psi^i(x)\}&=&N(x)\frac{\pi_{\psi^i}(x)}{\sqrt{|g|(x)}}+\xi^a(x)\psi^i_{,a}(x)\\
   \{H,\pi_{\psi^i}\}&=&\left(N(x)\sqrt{|g|(x)}g^{ab}(x)\right)_{,a}\psi^i(x)_{,b}+N(x)\sqrt{|g|(x)}g^{ab}(x)\psi^i(x)_{,a,b}+(\mathcal L_\xi \psi^i)(x).
 \end{array}
\end{equation}
To find the spacetime line element $ds^2=(g_{ab}\xi^a\xi^b-N^2)dt^2+2g_{ab}\xi^a dt dx^b+g_{ab}dx^adx^b$ at a point $(t_o,x_o^a)$ in the local chart we prepare at $t_o$ the first six components $(\psi^{ij},\pi_{\psi^{ij}})_{j\le i=1,2,3}$ of the scalar multiplet as
\begin{equation}
 \begin{array}{rcl}
  \psi^{ij}(x)&=&\epsilon \psi_o(x^i-x_o^i)(x^j-x_o^j)\\
  \pi_{\psi^{ij}}(x)&=&0,
 \end{array}
\end{equation}
where $\epsilon$ as well as the chart are small enough, so the test field approximation holds. We prepare the next three components $(\psi^{i},\pi_{\psi^{i}})_{i=1,2,3}$ as
\begin{equation}
 \begin{array}{rcl}
  \psi^{i}(x)&=&\epsilon \psi_o(x^i-x_o^i)\\
  \pi_{\psi^{i}}(x)&=&0,
 \end{array}  
\end{equation}
and one more $(\psi,\pi_{\psi})$ as
\begin{equation}
 \begin{array}{rcl}
  \psi(x)&=&0\\
  \pi_{\psi}(x)&=&\epsilon \pi_o.
 \end{array}
\end{equation}
The equations of motion yield the constituents $\left(\frac{N}{\sqrt{|g|}},\xi^a,N\sqrt{|g|}g_{ab}\right)$ at any desired $(x_o,t_o)$ through the observables
\begin{equation}
  \begin{array}{rcl}
    \dot \psi(x_o)&=&\epsilon N(x_o) \frac{\pi_o}{\sqrt{|g|(x_o)}}\\
    \dot \psi^i(x_o)&=&\epsilon \psi_o \xi^i(x_o)\\
    \dot \pi_{\psi^{ij}}(x_o)&=&2\epsilon N(x_o)\sqrt{|g|(x_o)}g^{ij}(x_o)\psi_o.
  \end{array}
\end{equation}
We now use the equivalence of the GR and SD equations of motion and can thus recover the spacetime line-element from physical observables in SD. 

\section{Conclusions}\label{sec:conclusions}

In this paper we showed two points about SD: 

First, the construction of SD as a theory that is equivalent to GR is not obstructed by standard matter couplings. In fact the construction of a linking theory that proves the equivalence of SD and GR needs only one additional ingredient: the conformal weight of the matter fields. We found that ``neutral coupling", i.e. giving all standard matter fields\footnote{In \cite{York3} a different scaling for matter fields was chosen, because of the use of non-canonical matter variables. However the presentation in \cite{IsenbergNester1,IsenbergNester2} suggests that ``neutral coupling" was discovered there.} conformal weight zero is a good, and perhaps the only,  choice for the coupling of standard matter\footnote{Notice that we use density weight $\frac 1 2$ spinors to ensure constraint decoupling.}. A matter that we did not discuss in this paper is that some models admit different conformal weights for matter fields in the construction of SD, but these will imply non-geometric generators of conformal transformations. I.e. the foliation in the equivalence to ADM is no longer CMC, but something else, which might furthermore be dependent on the gauge choice of the matter fields.  We thus find the fact the ``neutral coupling" works so well suggestive, because it leads to a notion of simultaneity that is completely described by the purely geometric constant mean curvature condition. We are thus inclined to propose neutral coupling as a canonical choice in the construction of SD and may elevate it to a construction principle. 

For matter with non-derivative coupling, using the neutral conformal coupling mentioned above we found a straightforward criterium  that either forbids or allows the coupling of a given type of field to SD. This is contained in equation \eqref{equ:inv_criterium}.  In this context we suggestively notice that the only term in the Standard Model Hamiltonian that imposes bounds on York time is the Higgs potential.  Outside of the Standard Model, but still within experimental status, the only other field that implies such a bound is the cosmological constant. It can be seen that de Sitter space does not obey the bound. The statement that the model does not obey the bound implies that the dual theory does not gauge fix the refoliations, and we are left with more than one ``global Hamiltonian" in the dual theory. One could wonder if this ambiguity is in fact not due to the high symmetry content of deSitter space. To this effect we note that we could always make the $\sigma^{ab}\sigma_{ab}$ term in \eqref{equ:inv_criterium} small enough so that no symmetry is present and yet it still breaks the bound. This raises interesting questions regarding the construction of a phenomenologically viable cosmological model within Shape Dynamics, a question which will be left for future work. 

Lastly, we found that an operational spacetime picture of SD can be obtained straightforwardly through a clock and rod model constructed from matter fields. At a superficial level this is surprising, because SD has a notion of simultaneity and naively it would seem that SD breaks Lorentz invariance. This would violate the very idea of spacetime that is at the heart of our understanding of special and general relativity. However, this conclusion is premature. Let us assume for a moment that we had access to GR-matter trajectories but not to the spacetime metric. Then we would have to adopt an operational description of spacetime much like the idea that underlies noncommutative geometry (see opening quote). We would then recover spacetime precisely along the lines outlined in section \ref{sec:spacetimeRecovery}. Hence the spacetime picture of SD is the same as an operationally defined spacetime picture in GR.

\section*{Acknowledgements}

HG thanks Sean Gryb for noticing that DeSitter doesn't satisfy the bound, and Steve Carlip for discussions on the physical meaning of this breaking. TK thanks UC Davis for hospitality. Research at the Perimeter Institute is supported in part by the Government of Canada through NSERC and by the Province of Ontario through MEDT.  HG was supported in part by the U.S.
Department of Energy under grant DE-FG02-91ER40674.

\end{document}